\documentclass[twocolumn,american]{revtex4-2}
\usepackage{mathptmx}

\usepackage[T1]{fontenc}
\usepackage[utf8]{inputenc}
\usepackage{babel}
\usepackage{array}
\usepackage{textcomp}
\usepackage{multirow}
\usepackage{amsbsy}
\usepackage{amstext}
\usepackage{graphicx}
\usepackage[unicode=true,pdfusetitle,
 bookmarks=true,bookmarksnumbered=false,bookmarksopen=false,
 breaklinks=false,pdfborder={0 0 1},backref=false,colorlinks=false]
 {hyperref}

\makeatletter

\newcommand{\lyxmathsym}[1]{\ifmmode\begingroup\def\b@ld{bold}
  \text{\ifx\math@version\b@ld\bfseries\fi#1}\endgroup\else#1\fi}

\providecommand{\tabularnewline}{\\}

\makeatother

\begin{document}
\title{Orbital localization and the role of the Fe and As $4p$ orbitals
in BaFe$_{2}$As$_{2}$ probed by XANES}
\author{A. G. de Figueiredo$^{1}$\emph{,} M. R. Cantarino$^{1}$\emph{, }W.
R. da Silva Neto$^{1,2}$, K. R. Pakuszewski$^{3}$, R. Grossi$^{3}$,
D. S. Christovam$^{3*}$, J. C. Souza$^{3}$, M. M. Piva$^{3*}$,
G. S. Freitas$^{3}$, P. G. Pagliuso$^{3}$, C. Adriano$^{3}$,\emph{
}F. A. Garcia$^{1}$}
\affiliation{$^{1}$Instituto de Física, Universidade de São Paulo, São Paulo-SP,
05508-090, Brazil}
\affiliation{$^{2}$Instituto de Química, Universidade de São Paulo, São Paulo-SP,
05508-090, Brazil}
\affiliation{$^{3}$Inst Fis Gleb Wataghin, Universidade Estadual de Campinas,
Campinas-SP, 13083-859, Brazil}
\altaffiliation{Current address: Max Planck Institute for Chemical Physics of Solids, Nöthnitzer Straße 40, 01187 Dresden, Germany}

\begin{abstract}
The polarization dependence of the near edge x-ray absorption spectroscopy (XANES) is an element specific probe to the real-space distribution of the density of unoccupied states in solid-state materials. In this paper, we present Fe and As $K$-edge experiments of Ba(Fe$_{1-x}$$M_{x}$)$_{2}$As$_{2}$ ($M=$ Mn, Co and $x=0.0$ and $0.08$). The experiments reveal a strong polarization dependence of the probed XANES spectra, which concerns mainly an increase of the intensity of electronic transitions when the beam polarization is set out of the sample's $ab$ crystallographic plane. The results show that states with $p_{z}$-orbital character dominate the density of unoccupied states close to the Fermi level. Partial substitution of Fe by Co is shown to decrease the intensity anisotropy, suggesting that Co promotes electronic transfer preferentially to states with $p_{z}$-orbital character. On the other hand, Mn substitution causes the increase of the spectra $p_{z}$-orbital anisotropy, which is proposed to take place by means of an enhanced local Fe $3d4p$ mixing, unveiling the role of Fe $4p$ states in the localization of the Fe $3d$ orbitals. Moreover, by comparing our results to previous experiments, we identify the relative mixing between Fe and the pnictide $4p_{x,y,z}$ orbitals as a clear divide between the electronic properties of iron arsenides and selenides. Our conclusions are supported by multiple-scattering theory calculations of the XANES spectra and by quantum chemistry calculations of Fe coordination electronic structure.
\end{abstract}
\maketitle

\section{Introduction}

Doping an electronic correlated phase often results in rich phase
diagrams containing regions dominated by strong electronic correlations
and other regions wherein Fermi liquid behavior is observed. The relatively
recently introduced Fe pnictide (FePns) materials \citep{kamihara_iron-based_2008}
denote a large family of high-temperature unconventional superconducting
materials that seems to lie on the border between weak and strong
electronic correlations \citep{hosono_iron-based_2015}. Indeed, the
description of the relevant electronic degrees of freedom in terms
of either localized or itinerant states underlines a major conceptual
divide in this field \citep{chubukov_pairing_2012,de_medici_selective_2014,yin_magnetism_2011,yin_kinetic_2011}.

The $122$ parent compound BaFe$_{2}$As$_{2}$ is a particularly
well explored FePn material. Phases derived from the partial substitution
of Fe by Co, Ba(Fe$_{1-x}$Co$_{x}$)$_{2}$As$_{2}$, is a much debated
subject mainly because it leads to a robust and clean superconducting
(SC) ground state \citep{sefat_superconductivity_2008,hosono_iron-based_2015}.
In the case of Ba(Fe$_{1-x}$Mn$_{x}$)$_{2}$As$_{2}$, however,
no SC phase is observed \citep{kim_electron-hole_2010}. In principle,
partial substitution at the Fe site by Mn pushes the BaFe$_{2}$As$_{2}$
electronic properties to a Mott region, whereas partial substitution
by Co tends to decrease its electronic correlations. Indeed, the endpoints
of the Ba(Fe$_{1-x}$$M$$_{x}$)$_{2}$As$_{2}$ ($M=$ Mn or Co)
materials, BaMn$_{2}$As$_{2}$ and BaCo$_{2}$As$_{2}$, are characterized
as an antiferromagnetic Mott insulator \citep{singh_magnetic_2009,singh_magnetic_2009-1}
and a weakly correlated metal \citep{sefat_renormalized_2009,anand_crystallography_2014},
respectively, somehow corroborating the proposed substitutional trends.

The BaFe$_{2}$As$_{2}$ structure features FeAs layers well spaced
by Ba$^{2+}$ cations, where Fe is fourfold coordinated by As atoms
arranged in a slightly distorted tetrahedral geometry. This structure
breaks the degeneracy of the Fe derived $3d$ states close to the
Fermi level, directly affecting the BaFe$_{2}$As$_{2}$ Fermi surface
composition, making the Fe coordination an important parameter to
understanding the $122$ materials electronic properties \citep{hosono_iron-based_2015}.
In this work, we focus precisely on the electronic properties of the
FeAs coordination complex, investigating the polarization and composition
dependencies of the near edge (XANES) Fe and As $K$-edges X-ray absorption
spectra of Ba(Fe$_{1-x}$$M$$_{x}$)$_{2}$As$_{2}$ ($M=$ Mn or
Co, $x=0.0$ and $x=0.08$) single crystals.

Hard X-ray absorption spectroscopy (XAS) experiments could provide
a handful of key information to the field of the FePns materials,
including the role of transition metal substitution on their electronic
\citep{bittar_co-substitution_2011,baledent_stability_2012,merz_electronic_2012,baledent_electronic_2015,yamaoka_electronic_2017,pelliciari_fluctuating_2021}
and structural properties \citep{granado_pressure_2011,cheng_charge_2015,chu_iron_2013,xu_local_2010,joseph_temperature-dependent_2016,hacisalihoglu_study_2016},
as well as the degree of electronic correlations \citep{pelliciari_fluctuating_2021,lafuerza_evidence_2017,chen_x-ray_2011,chen_x-ray_2011-1}
in these systems. Here, we take advantage of the dipole selection
rules dependence on the beam polarization that makes the XAS spectra
strongly angular dependent. Focusing on the XANES region of the XAS
spectra, we thus probe, in real space, the distribution of the density
of unoccupied electronic states in our materials.

For all investigated materials, our results suggest that the density
of unoccupied states at Fermi level has a predominant $p_{z}$ orbital
character, in agreement with the case of the iron arsenide SmFeAsO
\citep{chang_angular_2009}. Co substitution is shown to populate
preferentially the As $4p_{z}$ orbitals, characterizing a distinct
charge transfer anisotropy. Mn substitution increases the spectra
$p_{z}$ orbital anisotropy suggesting the localization of the Fe
$3d$ and A s$4p$ states, adding yet another piece of evidence that
the BaFe$_{2}$As$_{2}$ material can be tuned to a correlated electronic
phase \citep{lafuerza_evidence_2017}, but this time not by doping.
From our calculations and experiments, we conclude that the BaFe$_{2}$As$_{2}$
electronic structure displays a delicate interplay between the local
Fe $3d4p$ mixing and the metal-to-ligand Fe $3d$ As$4p$ mixing.
Our findings unveil the role of the Fe $4p$ states and allow one
to make a clear distinction between the iron arsenides and the iron
selenides, for which the unoccupied states were found to have a predominant
Se $4p_{x,y}$ character \citep{joseph_study_2010}.

Our experiments are analyzed based on two complementary frameworks.
We first present multiple scattering theory calculations (as implemented
by the FEFF$8.4$ code \citep{zabinsky_multiple-scattering_1995,rehr_theoretical_2000}),
where we found relative agreement between our calculations and experiments
for the Fe $K$-edge and As $K$-edge. Nevertheless, FEFF is not able
to capture the angular dependence of the probed spectra. Then, we
present multiconfigurational electronic structure calculations of
the orbital states of a single FeAs$_{4}$ coordination complex, as
implemented by ORCA $5.0$ \citep{neese_orca_2012,neese_software_2018}.

\section{Material and Methods}

Ba(Fe$_{1-x}$$M$$_{x}$)$_{2}$As$_{2}$ ($M=$ Mn or Co, $x=0.0$
or $0.08$) single crystals were synthesized by an In-flux technique
as described in Ref. \citep{garitezi_synthesis_2013}. The Laue patterns
of all samples were obtained to determine the relative orientation
of the $\boldsymbol{a}$ and $\boldsymbol{b}$ crystallographic axis.
A composition ($x$) vs. temperature ($T$) phase diagram of Mn and
Co substituted samples is presented in Fig. \ref{fig:oppeningFigure}$(a)$
for context. Black squares and blue circles mark the data from Ref.
\citep{thaler_physical_2011} and red stars mark the data for our
samples, as determined from resistivity measurements \citep{garitezi_synthesis_2013,garcia_anisotropic_2019,rosa_possible_2014}.
The investigated samples are indicated by black arrows.

The Fe $K$-edge and As $K$-edge hard X-ray absorption spectroscopy
experiments were performed at the XDS beamline \citep{lima_xds_2016}
of the Brazilian Synchrotron Light source (CNPEM-LNLS) at room temperature.
Data was acquired by the partial Fe and As fluorescence. The samples
were mounted in a Huber $6+2$-circle diffractometer with the experimental
geometry as explained in Fig. \ref{fig:oppeningFigure}$(b)$. The
beam polarization direction is fixed and is aligned to one of the
sample crystallographic axes, which will be hereafter called the $\boldsymbol{a}$
axis. XANES spectra were thus collected for rotations about the in
plane $\boldsymbol{a}$ and $\boldsymbol{b}$ axis and the out of
plane $\boldsymbol{c}$ axis which are, respectively, termed $\theta$,
$\phi$ and $\chi$ rotations (see Fig. \ref{fig:oppeningFigure}$(b)$).
For all rotations, we adopted an angular ($\alpha$) interval of $0<\alpha<45\lyxmathsym{\textdegree}$.
All spectra were normalized in ATHENA \citep{ravel_athena_2005}.

FEFF calculations \citep{rehr_theoretical_2000} were adopted to calculate
the XANES spectra. The\emph{ ab initio} calculations were performed
and converged for clusters of up to $282$ atoms for the Fe $K$-edge
and $144$ atoms for the As $K$-edge. In both cases, the Hedin-Lundqvist
\citep{hedin_explicit_1971} pseudopotential was adopted to account
for the effect of the local exchange-correlation. Self-consistent
calculations were performed for a cluster radius of $6.5$ $\text{Å}$
(Fe $K$-edge) and $8.0$ $\text{Å}$ (As $K$-edge). Quadrupolar
transitions and spin orbit coupling effects were considered but nos
sizable effects were observed. Chemical substitution effects were
simulated in the case of the As $K$-edge, by substituting one in
ten Fe atoms by one Mn (Co) atom. The large self-consistent $8$$\text{Å}$
radius was adopted to include all dopants in the self-consistent calculations.

Multiconfigurational electronic calculations of the electronic states
of a FeAs$_{4}$ tetrahedral like molecule were implemented by ORCA
$5.0$ \citep{neese_orca_2012,neese_software_2018}. Ionic charges
were set as $-3$ for each arsenic ion and $+2$ for the metallic
center, resulting in a $S=2$ molecule, consistent with a high spin
complex within tetrahedral like coordination ($D_{2d}$ symmetry).
Atomic positions were obtained from the BaFe$_{2}$As$_{2}$ crystallographic
data.

To capture the optimized configuration of the electronic states in
this $3d$ metal complex, we apply the CAS-SCF method and electronic
correlations were later taken into account by NEVPT-2 calculations.
The relativistic adapted Karlsruhe valence triple-zeta with two sets
of polarization functions were adopted as the basis set within the
ZORA approximation \citep{singh_covalency_2017}. We built a complete
active space (CAS) consisting only of the five $3d$ orbitals and
its six electrons - CAS(6,5) - in the $D_{2d}$ symmetry, and the
ligand Field Parameters were obtained at the end of the calculation.
Virtual states, $5$eV above the Fermi level, were calculated similarly
but adopting a CAS(2,12) \citep{norman_simulating_2018}.

\section{Results and Discussion}

\begin{figure*}
\begin{centering}
\includegraphics[width=1\textwidth]{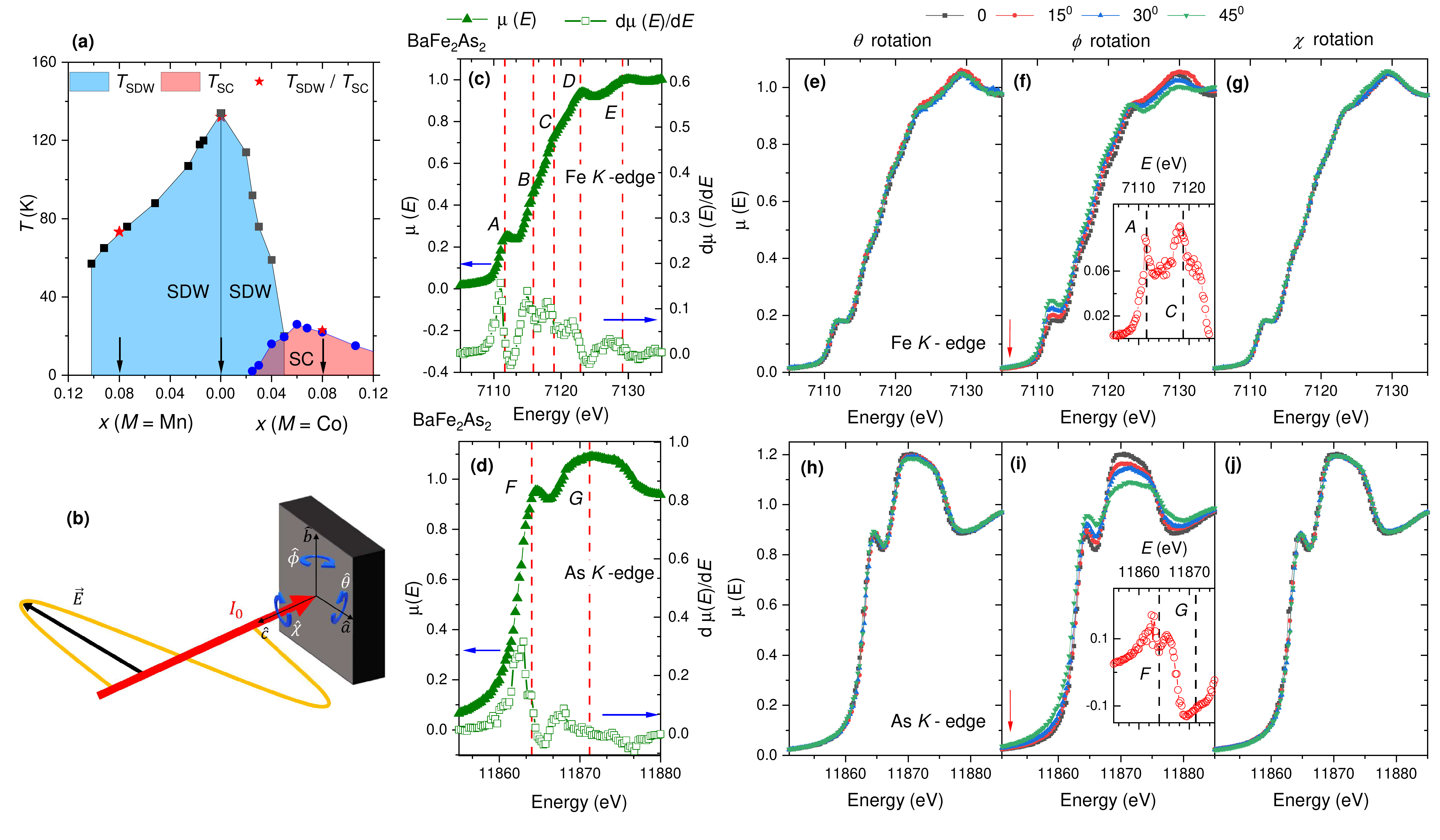}
\par\end{centering}
\caption{$(a)$ Composition ($x$) $vs.$ $T$ phase diagram for the Ba(Fe$_{1-x}$Mn$_{x}$)$_{2}$As$_{2}$
and Ba(Fe$_{1-x}$Co$_{x}$)$_{2}$As$_{2}$ transition metal substituted
iron arsenides (black squares are data from Ref. \citep{thaler_physical_2011}).
$(b)$ Schematic representation of the experimental geometry, defining
the rotation angles $\phi$, $\theta$ and $\chi$ as rotations around
the $\boldsymbol{a}$, $\boldsymbol{b}$ and $\boldsymbol{c}$ axis,
respectively. $(c)$ Representative BaFe$_{2}$As$_{2}$ Fe $K$-edge
normalized XANES spectrum ($\mu(E)$, left axis, full symbols) and
its derivative ($d\mu(E)/dE$, right axis, open symbols). Capital
letters $A$-$E$ mark the transitions and the dashed lines associate
the features in the spectrum derivative to the spectrum. $(d)$ The
respective As $K$-edge data of BaFe$_{2}$As$_{2}$, with capital
letters $F$-$G$ labeling the absorption features $(e)-(j)$. Polarization
dependence of the Fe $K$-edge and As $K$-edge XANES spectra of BaFe$_{2}$As$_{2}$.
The open red circles in the insets of panels $(f)$ and $(i)$ show
the difference spectrum ($\mu(E)_{\phi=45}-\mu(E)_{\phi=0}$) with
the dashed lines marking some representative energy positions. \label{fig:oppeningFigure}}
\end{figure*}

In Figs.\ref{fig:oppeningFigure} $(c)$ and $(d)$ we present representative
normalized ($\mu(E)$) Fe and As $K$edge spectra of BaFe$_{2}$As$_{2}$
along with their respective spectra derivatives. The Fe $K$-edge
XANES spectrum presents five absorption features labeled by capital
letters $A-E$. The features are positioned in the mid range between
the maxima and minima of the spectrum derivative. Features $A-C$
are of electronic nature, whereas features $D$ and $E$ are predominantly
due to scattering processes \citep{bittar_co-substitution_2011,merz_electronic_2012,baledent_stability_2012}.

The $A$ feature sits at about $E_{\text{F}}$, which is found to
be $E_{\text{F}}\approx7111.6$ eV. This feature is called the pre-edge
and is understood to stem from a dipolar transition from the Fe $1s$
to Fe$3d$As$4p$ hybrid bands. It comprises a series of energy levels
located in a $1-2$ eV bandwidth about $E_{\text{F}}$ and expresses
the properties of the FeAs tetrahedral coordination complex. Feature
$B$ denote transitions to hybrid states (including Fe$4s$ and As$4p$
states) lying $3$ eV above the Fe$3d$As$4p$ states. Feature $C$,
at about $\approx7118.6$ eV, is the main atomic transition of the
Fe $K$-edge and stems from Fe $1s\rightarrow4p$ transitions \citep{groot_1s_2009,vanko_intersite_2008}.

The As $K$-edge XANES spectrum in Fig.\ref{fig:oppeningFigure} $(d)$
display two main features, which we called $F$ and $G$. The former
identify the main As $1s\rightarrow4p$ atomic electronic transition
which, due to the As coordination, also includes contributions from
Fe $3d$ orbitals. The latter is mainly due to scattering processes
and transitions to excited electronic states $5$ eV above $E_{\text{F}}$.
The $F$ feature sits about $E_{F}\approx11864.3$ eV.

In Fig.\ref{fig:oppeningFigure} $(e)-(g)$ we present the Fe $K$-edge
spectra of BaFe$_{2}$As$_{2}$ for all investigated rotations. In
its local coordinate frame, rotating the sample is equivalent to changing
the incident beam polarization, which leads to new selection rules
for the dipole transitions. A $\theta$ rotation is thus a control
experiment, which does not change the beam polarization. Indeed, a
direct inspection Fig.\ref{fig:oppeningFigure}$(e)$ reveal that
$\theta$ rotations do not change the spectra. A $\chi$ rotation
probes orbitals with planar components (as the $p_{x}$ and $p_{y}$
$p$-orbitals ) whereas $\phi$ rotations probe orbitals with $z$
symmetry, as $p_{z}$ orbitals.

The pre-edge intensities of the observed spectra clearly increase
under $\phi$ rotations (Fig.\ref{fig:oppeningFigure}$(f)$) characterizing
the spectra anisotropy. The inset in Fig.\ref{fig:oppeningFigure}$(f)$
display the difference spectrum, obtained from making $\mu(E,\phi=45\lyxmathsym{\textdegree})-\mu(E,\phi=0)$.
As it is clear, the anisotropy is strong in the pre-edge ($A$ feature)
but persists in all the region of the electronic transitions, peaking
again close to the main edge ($C$ feature). The red arrow in the
figure calls attention to the fact that the baseline of the spectra
coincides in the region below the $E_{\text{F}}$, excluding a systematic
shift of the background signal as a source of the effect. Our analysis
will focus on the electronic transitions and in particular in the
pre-edge transition.

The significant increase in the pre-edge intensity for a $\phi$ rotation
shows that the mixing between Fe$3d$ and As $4p_{z}$ orbitals form
states with a higher density of unoccupied states than the Fe$3d$As$4p_{x,y}$
orbitals. The observed lack of in-plane anisotropy for a $\chi$ rotation
is expected because the $p_{x}$ and $p_{y}$ orbital symmetry is
not reduced in a tetragonal environment \citep{stohr_magnetism_2006}.
Indeed, it is to be noted that the Fe derived $3d$ states are observed
via their hybridization with $p$ states and, therefore, the $p$
states symmetries are the relevant properties in discussing the pre-edge
polarization dependence. In addition, the very observation of the
XAS polarization dependence is unexpected since in itinerant electron
systems the ligands are expected to be weak due to screening by conduction
electrons. These results add to the importance of the local electronic
properties of itinerant magnets \citep{mounssef_hard_2019}.

In Figs.\ref{fig:oppeningFigure} $(h)-(j)$, we show the As $K$-edge
spectra data. Here, one can also observe that $\theta$ and $\chi$
rotations do not change the spectra. Again, the spectra are clearly
anisotropic for $\phi$ rotations, with the absorption edge becoming
more intense. This result provides a direct assessment of the As $4p_{z}$
relative lower electronic filling and larger anisotropic orbital character.
We also call attention to the red arrow in \ref{fig:oppeningFigure}
$(i)$, showing that the baseline of the spectra coincides in the
region below the Fermi level. In the inset, we show the difference
spectrum which evidence the large anisotropy of both $F$ and $G$
features.

\begin{figure}
\begin{centering}
\includegraphics[width=1\columnwidth]{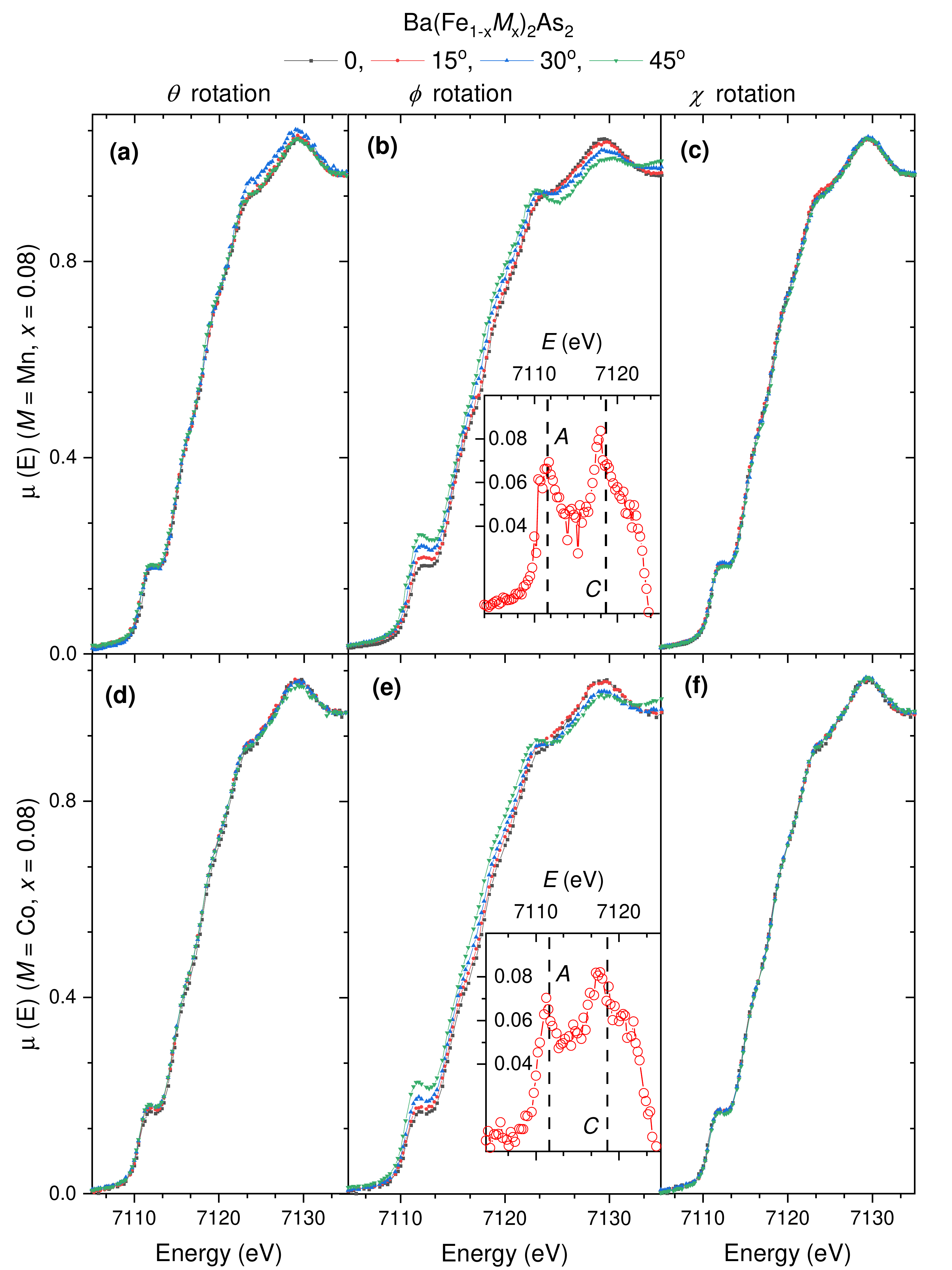}
\par\end{centering}
\caption{Fe $K$-edge XANES spectra of $(a)-(c)$ Ba(Fe$_{1.92}$Mn$_{0.08}$)$_{2}$As$_{2}$
and $(d)-(f)$ Ba(Fe$_{1.92}$Co$_{0.08}$)$_{2}$As$_{2}$ as a function
of the polarization (see figures' top). In all cases, the normalized
intensities ($\mu(E))$ are presented. The open red circles in the
insets of panels $(b)$ and $(e)$ show the difference spectrum ($\mu(E)_{\phi=45}-\mu(E)_{\phi=0}$)
with the dashed lines marking some representative energy positions.
\label{fig:FeKedge}}
\end{figure}

Our next step is to investigate the composition dependence of the
above effects. We start by inspecting the Fe $K$-edge of Mn and Co
substituted samples. Their XANES normalized spectra ($\mu(E)$) are
presented in figures \ref{fig:FeKedge}$(a)-(f)$. The Co substituted
sample is a superconductor with $T_{SC}\approx22$ K whereas the Mn
rich sample does not display SC (see Fig. \ref{fig:oppeningFigure}$(a)$).
The putative electronic effects of the Co and Mn substitutions would
be symmetric with respect to hole and electron doping making these
samples ideal for our studies. Concerning $\theta$ and $\chi$ rotations,
the results are similar to what was found in the parent compound,
whereas the data for $\phi$ rotations suggest a weak composition
dependency of this anisotropy.

The insets in figures \ref{fig:FeKedge}$(b)$ and $(e)$ present
the difference spectra for the Mn and Co substituted samples, respectively,
and show this effect in more detail. Moreover, the $C$ feature remains
markedly anisotropic under $\phi$ rotations, showing that Fe$4p_{z}$
orbitals form bands with a higher unoccupied density of states than
the Fe$4p_{x,y}$ orbitals. This particular result for the BaFe$_{2}$As$_{2}$
parent compound and doped materials is in contrast to the case of
SmFeAsO \citep{chang_angular_2009}. In addition, as in previous XANES
experiments of iron arsenides, the $C$ feature is not much affected
by Co substitution \citep{bittar_co-substitution_2011,baledent_electronic_2015,baledent_stability_2012,yamaoka_electronic_2017}
and is here shown to be unaffected by Mn substitution as well.

\begin{figure}
\begin{centering}
\includegraphics[width=1\columnwidth]{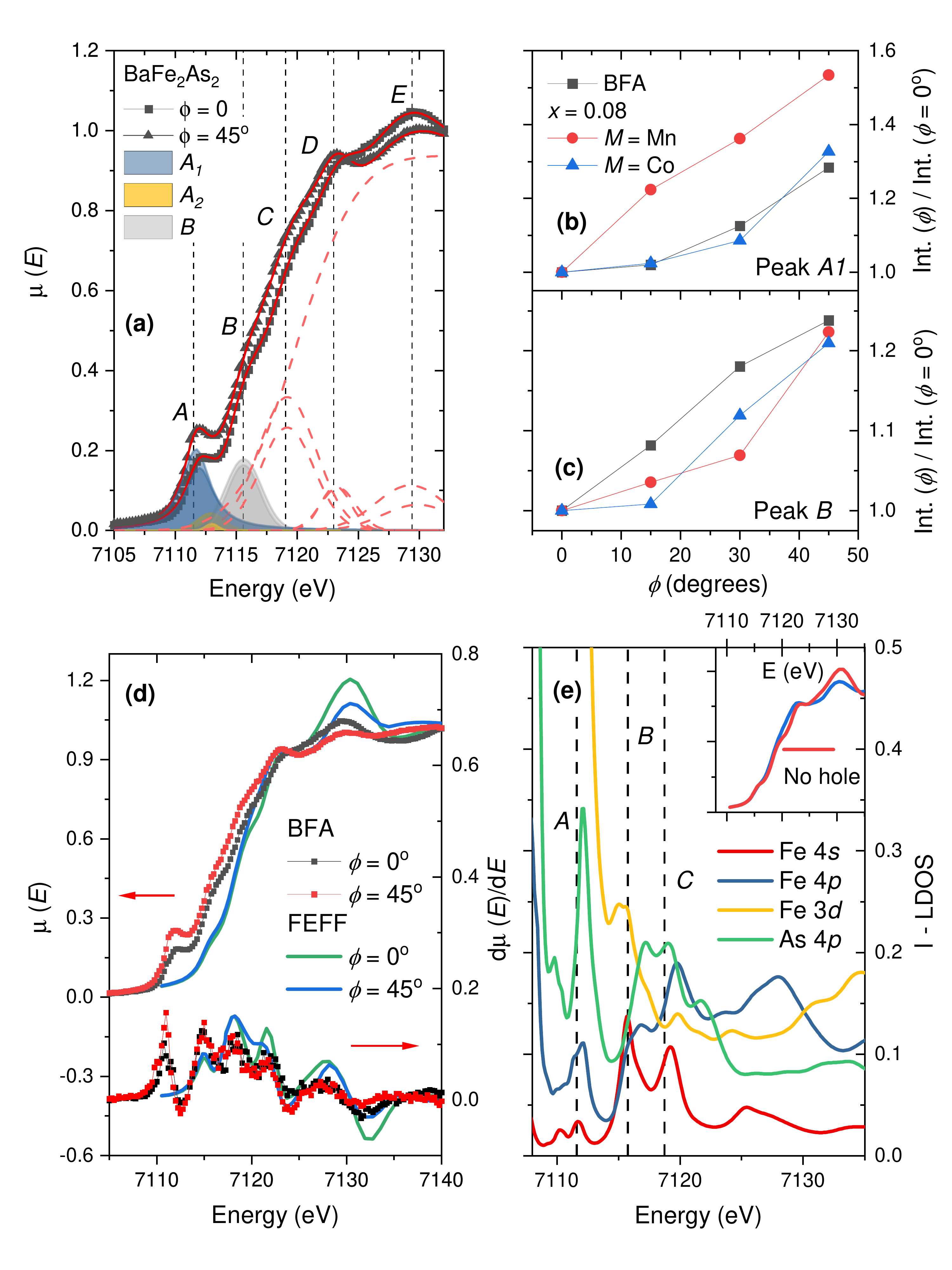}
\par\end{centering}
\caption{$(a)$ Fe $K$- edge XANES spectra of BaFe$_{2}$As$_{2}$ ($\phi=0$
and $\phi=45\lyxmathsym{\protect\textdegree})$ and their respective
phenomenological fittings. As shown, two peaks, termed $A_{1}$ and
$A_{2}$, are included to describe the pre-edge $A$ feature. The
shaded regions below the curves are the peak areas which are adopted
as the resonance intensities. The obtained intensities are presented
in $(b)$ and $(c)$ for, respectively, the $A_{1}$ pre-edge peak
and the $B$ edge peak as a function of $\phi$ and composition. In
$(d)$ we present polarization-dependent FEFF calculations of the
Fe $K$-edge spectra (left axis) and its derivative (right axis) compared
the respective experimental data. A shift of $-1.3$ eV and a broadening
of $1$ eV were considered in the calculations. In panel $(e)$, we
show the FEFF calculated site and orbital projected LDOS of BaFe$_{2}$As$_{2}$.
Features $A$, $B$ and $C$ are presented for comparison to the results.
The inset of the figure compares the FEFF calculated spectra ($\phi=45\lyxmathsym{\protect\textdegree}$)
with and without core-hole effects. \label{fig:FeKedgeplus}}
\end{figure}

The composition effect is better observed by our analysis in figures
\ref{fig:FeKedgeplus}$(a)-(c)$. To quantify the intensities anisotropy
as a function of $\phi$, we fit the Fe $K$-edge spectra by a Lorentzian
(the $A_{1}$ peaks) and Gaussian (the $A_{2}$, $B$, $C$, $D$
and $E$ peaks) lineshapes plus a Fermi-Dirac function as in figure
\ref{fig:FeKedgeplus}$(a)$. The two peaks in the pre-edge region
are clearly present in high-resolution experiments \citep{yamaoka_electronic_2017}
and are suggested by the analysis of our spectra derivatives. Indeed,
adopting two peaks is instrumental to extract a consistent fitting
analysis. The resonance intensities $I_{\text{peak}}$ are estimated
from the peak areas, as in the shades of figure \ref{fig:FeKedgeplus}$(a)$.

The $A_{1}$ feature sits just at $E_{F}$ whereas the $B$ feature
lies $\approx3$ eV above it. We thus compare the chemical substitution
electronic effects in these two distinct situations by tracking the
anisotropy behavior of the $A_{1}$ and $B$ features. In figures
\ref{fig:FeKedgeplus}$(b)-(c)$, respectively, we plot the $I_{A_{1}}$
and $I_{B}$ normalized intensities as functions of $\phi$ for the
compositions as indicated. Each data set is normalized as $I_{\text{peak}}(\phi)/I_{\text{peak}}(\phi=0)$.

The $A_{1}$ peak of the Mn substituted sample is distinctly more
anisotropic, but the composition effect is not present in the case
of the $B$ feature. Since the latter lies $\approx3$ eV above $E_{\text{F}}$,
it would be hardly affected by chemical substitution, as observed.
A naive interpretation about the effect of Mn substitution is that
Mn ``doping'' fills the Fe$3d$As$4p$ hybrid bands with holes,
increasing the amount of unoccupied states. At the present ``doping''
level, however, Mn impurities do not act as charge dopants to BaFe$_{2}$As$_{2}$
\citep{suzuki_absence_2013,texier_mn_2012} and we shall return to
this discussion later by proposing a distinct mechanism for the effects
of chemical substitution.

In many instances \citep{zabinsky_multiple-scattering_1995,ankudinov_real-space_1998,rehr_theoretical_2000},
FEFF calculations provide a first approach to the interpretation of
the XANES spectra. In figure \ref{fig:FeKedgeplus}$(d)$ we show
polarization dependent FEFF calculations of the BaFe$_{2}$As$_{2}$
Fe $K$- edge spectra, and their derivatives, compared to experimental
data. The calculations reproduce well the $B-E$ features position
but their polarization dependencies are not fully reproduced. Moreover,
the $A$ feature is missed completely, reflecting that FEFF calculations
do not capture in full the properties of bound states \citep{ankudinov_real-space_1998,zabinsky_multiple-scattering_1995,rehr_theoretical_2000}. 

If core-hole effects can be considered weak, the nature of the observed
transitions can be associated with the element (site) and orbital
projected local density of states (LDOS). In figure \ref{fig:FeKedgeplus}$(e)$,
we present the LDOS obtained from FEFF calculations, focusing on the
Fe and As derived states. High densities of states are predicted at
the positions of the $A$, $B$ and $C$ features. In particular,
the $B$ and $C$ features correlate, respectively, to the LDOS due
to the Fe $4s$ and As and Fe $4p$ states. As anticipated in our
discussion, the $A$ feature can be associated with a high LDOS derived
from Fe $3d$ and As $4p$ states. In addition, our calculations also
predict a high density of Fe $4p$ states about this same region,
inviting an investigation into the role of the local Fe$3d4p$ hybridization,
which we shall discuss based on quantum chemistry calculations. The
inset of figure \ref{fig:FeKedgeplus}$(e)$, compares the FEFF calculated
spectra with and without core-hole effects and the close similarity
between the calculations suggests that our discussion is adequate
in a first approximation.

\begin{figure}
\begin{centering}
\includegraphics[width=1\columnwidth]{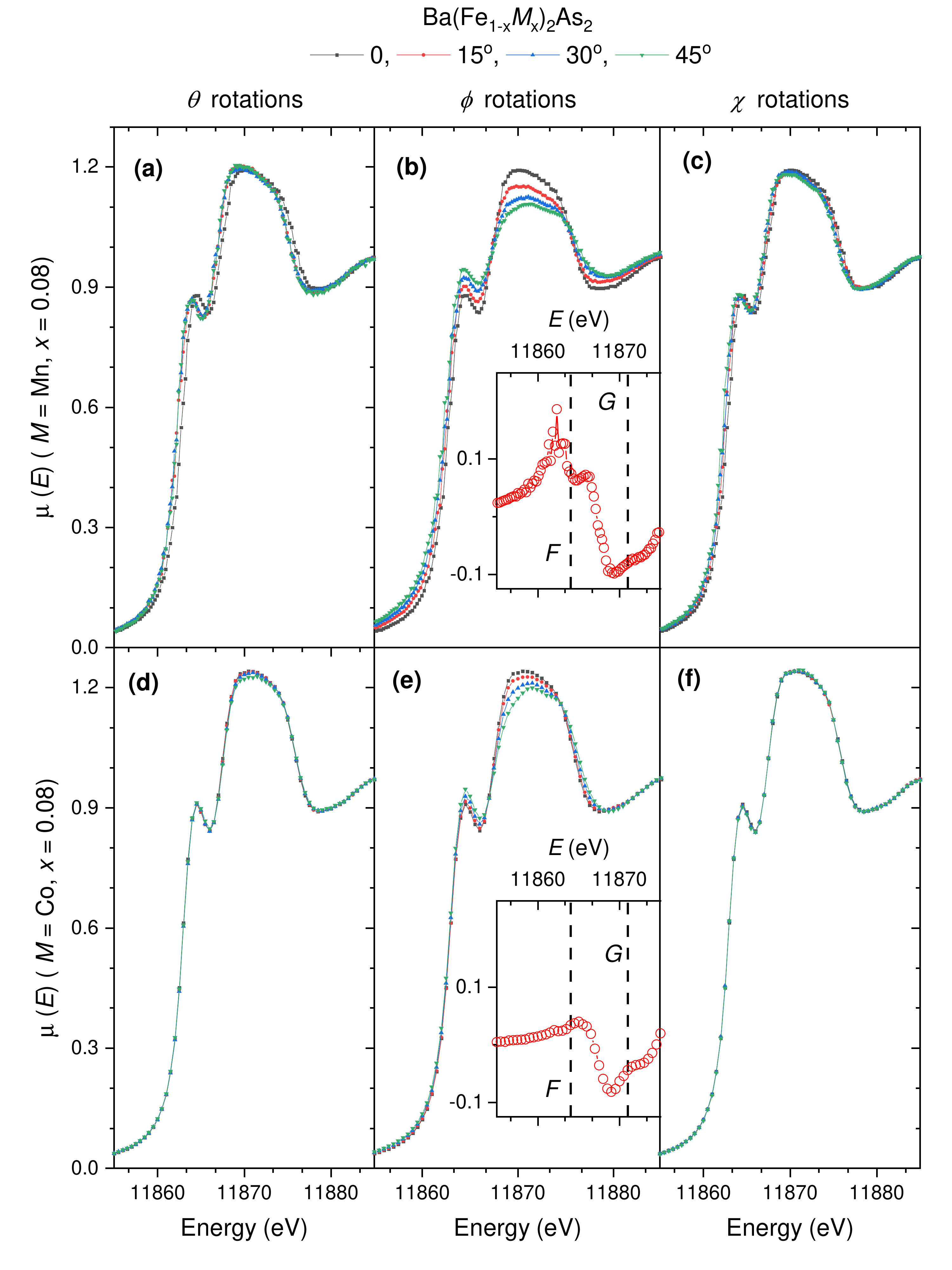}
\par\end{centering}
\caption{As $K$-edge XANES spectra of $(a)-(c)$ Ba(Fe$_{1.92}$Mn$_{0.08}$)$_{2}$As$_{2}$
and $(d)-(f)$ Ba(Fe$_{1.92}$Co$_{0.08}$)$_{2}$As$_{2}$ as a function
of the polarization (see figures' top). In all cases, the normalized
intensities ($\mu(E))$ are presented. The open red circles in the
insets of panels $(b)$ and $(e)$ show the difference spectrum ($\mu(E)_{\phi=45}-\mu(E)_{\phi=0}$)
with the dashed lines marking some representative energy positions.
\label{fig:AsKedge }}
\end{figure}

We now turn to the As $K$-edge experiments of other compositions.
In figures \ref{fig:AsKedge }$(a)-(f)$, the normalized intensities
($\mu(E)$) of the As $K$-edge XANES spectra of the doped samples
are presented. Again, the measured spectra are isotropic under $\theta$
and $\chi$ rotations. In the case of $\phi$ rotations, however,
both Ba(Fe$_{1.92}$Mn$_{0.08}$)$_{2}$As$_{2}$ and Ba(Fe$_{1.92}$Co$_{0.08}$)$_{2}$As$_{2}$
spectra are polarization-dependent but this time the composition effect
is straightforwardly observed by direct inspection of figures \ref{fig:AsKedge }$(b)$
and \ref{fig:AsKedge }$(e)$. The insets of the same figures present
the respective difference spectrum ($\mu(E)_{\phi=45}-\mu(E)_{\phi=0}$).
The insets are on the same scale, making it clear that the XANES anisotropy
is larger for the Mn rich material. Moreover, it is also clear that
the edge anisotropy of the Co substituted sample decreases when compared
to the case of the parent compound, whereas it increases for the Mn
containing sample. Since the As $K$-edge is a direct probe to the
properties of the As $4p_{x,y,z}$ orbitals sitting at about $E_{\text{F}}$,
the distinction between the localization and occupation of the As
$4p_{z}$ orbitals should be more evident, as observed.

The anisotropy of the post edge feature at about $11870$ eV, which
is about $5$ eV above the Fermi level, is also affected by composition,
being less intense for the Co substituted sample. Since this feature
is well above the Fermi level, the composition effect is most likely
due to the direct effect of the impurity scattering potential \citep{kobayashi_carrier_2016,merz_substitution_2016}
and we conclude that Mn impurities act as stronger scattering centers
than Co.

\begin{figure}
\begin{centering}
\includegraphics[width=1\columnwidth]{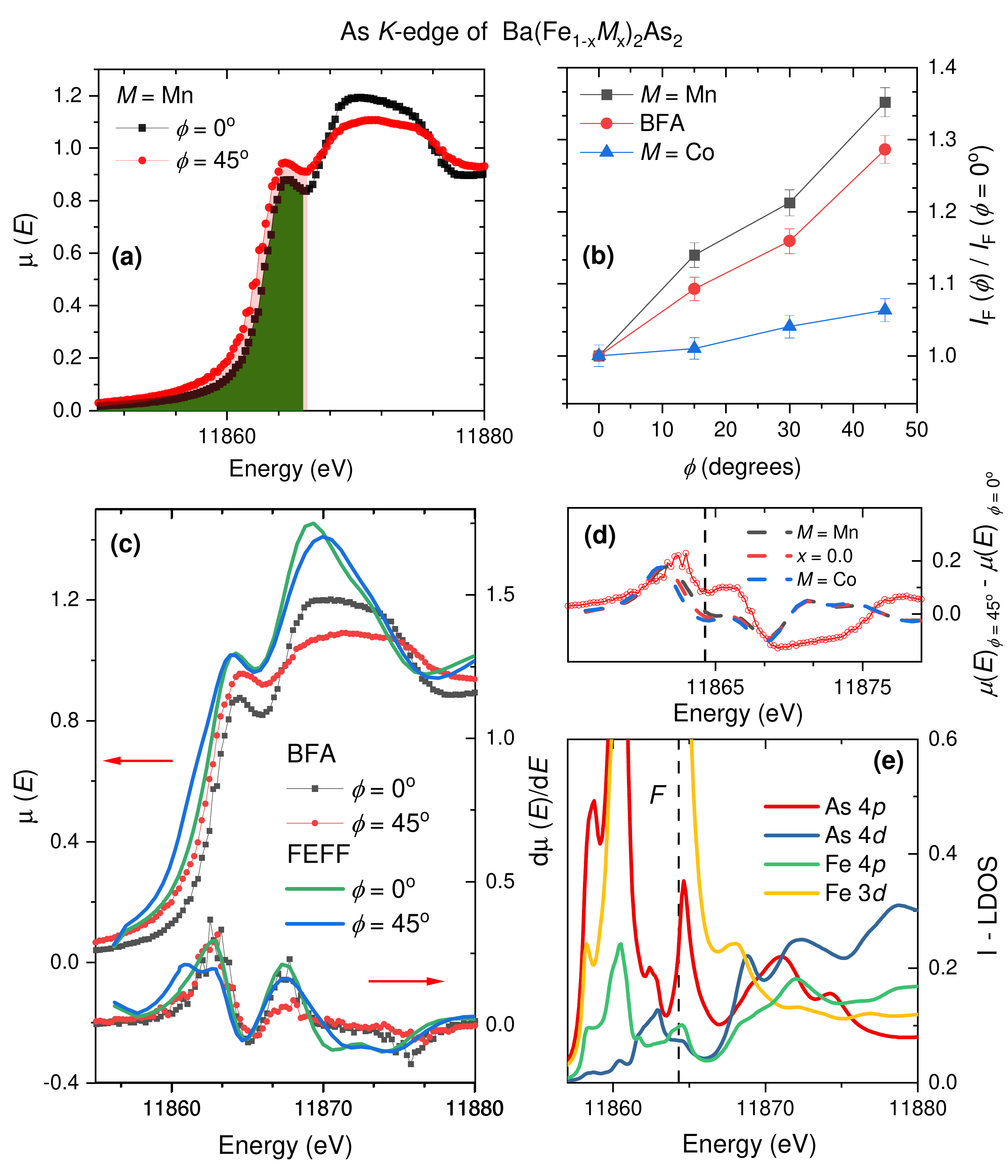}
\par\end{centering}
\caption{$(a)$ Normalized intensities ($\mu(E)$) of the As $K$-edge XANES
spectra of Ba(Fe$_{1.92}$Mn$_{0.08}$)$_{2}$As$_{2}$ for $\phi=0$
and $\phi=45\lyxmathsym{\protect\textdegree}$. In each case, the
panel shows the spectrum region taken into consideration to perform
the numerical integration that represents the $F$ feature intensity
($I_{F}(\phi)$). $(b)$ The polarization dependence of the $F$ peak
intensity as a function of $\phi$, for all compositions. In $(c)$
we present polarization-dependent FEFF calculations of the As $K$-edge
spectra (left axis) and its derivative (right axis) compared the respective
experimental data. A shift of $1$ eV and a broadening of $1.2$ eV
were considered in the calculations. In panel $(d)$, we present the
difference spectrum of the BaFe$_{2}$As$_{2}$ data (open red circles)
compared to FEFF calculations of Ba(Fe$_{1-x}$$M_{x}$)$_{2}$As$_{2}$
, with $M=$ Mn or Co and $x=0.1$ (see main text). $(e)$ FEFF calculated
site and orbital projected LDOS of BaFe$_{2}$As$_{2}$ focusing on
the Fe and As derived electronic states. \label{fig:AsKedgeplus}}
\end{figure}

To estimate the composition effect in the edge intensity anisotropy
as a function of $\phi$ {[}$I(\phi)${]}, we adopt again the integrated
areas of the edge features as the approximation to $I(\phi)$. This
time, we undertake a direct approach numerically integrating the experimental
data, as exemplified in figure \ref{fig:AsKedgeplus}$(a)$. Indeed,
results from a lineshape analysis of the As $K$-edge are too much
dependent on the choice of parameters and we found that a numerical
integration suffices to properly capture our results. For each $\phi$,
$I(\phi)$ is obtained and then the data is normalized as $I(\phi)/I(\phi=0)$.
Results for all samples are shown in figure \ref{fig:AsKedgeplus}$(b)$.
The error bars are estimated by small variations of the integration
region. In comparison to the parent compound, the effect of Co substitution
is clear, whereas that of Mn is weak but significant.

In Ref. \citep{baledent_electronic_2015}, Co substitution is suggested
to populate the As derived orbitals. Here, we show that the electrons
derived from Co substitution go preferentially to the unoccupied As$4p_{z}$
states, characterizing a distinct charge transfer anisotropy. By causing
an electronic transfer to orbitals along the $c$-axis, Co substitution
unbalances the electron and holes contributions to the transport along
this direction, increasing the incoherent scattering in this direction.
We thus support the interpretation given in Ref. \citep{nakajima_comprehensive_2018}
to the observed BaFe$_{2}$As$_{2}$ inter plane resistivity anisotropy
and its increase with Co substitution \citep{tanatar_systematics_2011,nakajima_comprehensive_2018}.
Moreover, our results provide a real space picture of the evolving
$3$D character of the BaFe$_{2}$As$_{2}$ electronic structure as
previously probed by angle-resolved photoemission spectroscopy (ARPES)
\citep{thirupathaiah_orbital_2010}.

On the other hand, no change in electronic filling can be invoked
as a mechanism to the anisotropy increase caused by Mn substitution.
As we will discuss, this effect is connected to the localization of
the Fe-derived electronic states. Indeed, isoelectroctronic substitutions,
as observed by resonant inelastic x-ray scattering experiments of
Mn- and P-substituted BaFe$_{2}$As$_{2}$ \citep{garcia_anisotropic_2019,pelliciari_reciprocity_2019},
may increase electronic correlations.

We performed polarization-dependent FEFF calculations, aiming at describing
two effects: the spectra anisotropy and the composition dependence
of this effect. In figure \ref{fig:AsKedgeplus}$(c)$, we compare
the experimental and calculated As $K$-edges of BaFe$_{2}$As$_{2}$
for $\phi=0$ and $\phi=45\text{°}$ and their respective derivatives.
The overall spectral shape and anisotropy are well reproduced but
there is a lack of detail in the effect of the anisotropy. In the
calculations, the edge peaks nearly coincide whereas in the experiments
the edge intensity of the $\phi=45\text{°}$ spectra sits above the
$\phi=0$ spectra for all compositions.

To simulate the effects of chemical substitution, we performed FEFF
calculations replacing one in ten Fe atoms with a dopant (either Mn
or Co). This is equivalent to a $x=0.1$ composition, which is close
to our samples for which $x=0.08$. Three dopant distributions were
calculated and then averaged out. In figure \ref{fig:AsKedgeplus}$(d)$,
we compare the difference spectrum ($\mu(E)_{\phi=45}-\mu(E)_{\phi=0}$)
obtained from the BaFe$_{2}$As$_{2}$ data (open red circles) and
from FEFF calculations of the parent compound and substituted samples
(thick lines). As can be observed, the edge polarization dependence
is partially reproduced but only a small composition effect is observed
in the FEFF calculations.

In figure \ref{fig:AsKedgeplus}$(e)$, we show the site and orbital
projected LDOS obtained from FEFF calculations. The $F$ feature position
is marked for comparison. It shows that the main edge is dominated
by As $4p$ states, as expected. Moreover, the high density of Fe
$3d$ states about the edge position allows the formation of Fe$3d$As$4p$
hybrid bands making the As $K$-edge transition sensitive to this
mixing. The LDOS of the Fe $4p$ derived states also peaks about the
$F$, as also calculated in the case of the Fe $K$-edge, further
suggesting its relevance to the electronic states about the Fermi
energy. Here, due to the As $K$-edge higher energy, core-hole effects
are likely less relevant than for the Fe $K$-edge, (see inset of
figure \ref{fig:FeKedgeplus}$(e)$), validating the present discussion
in a first approximation.

Whereas our FEFF calculations for the Fe and As K edges are able to
support our discussion of the experimental results, it is clear that
some aspects of the physics of our system are not captured. This is
nothing but an expression of the still unresolved character of the
electronic states of the FePns materials, that lay in the border between
strong and weak electronic correlations. We thus resort to quantum
chemistry calculations of the electronic properties of a single FeAs$_{4}$
distorted tetrahedron, within a multiconfigurational calculation of
the orbital states \citep{neese_orca_2012,neese_software_2018,norman_simulating_2018,singh_covalency_2017}.

\begin{figure}
\begin{centering}
\includegraphics[width=1\columnwidth]{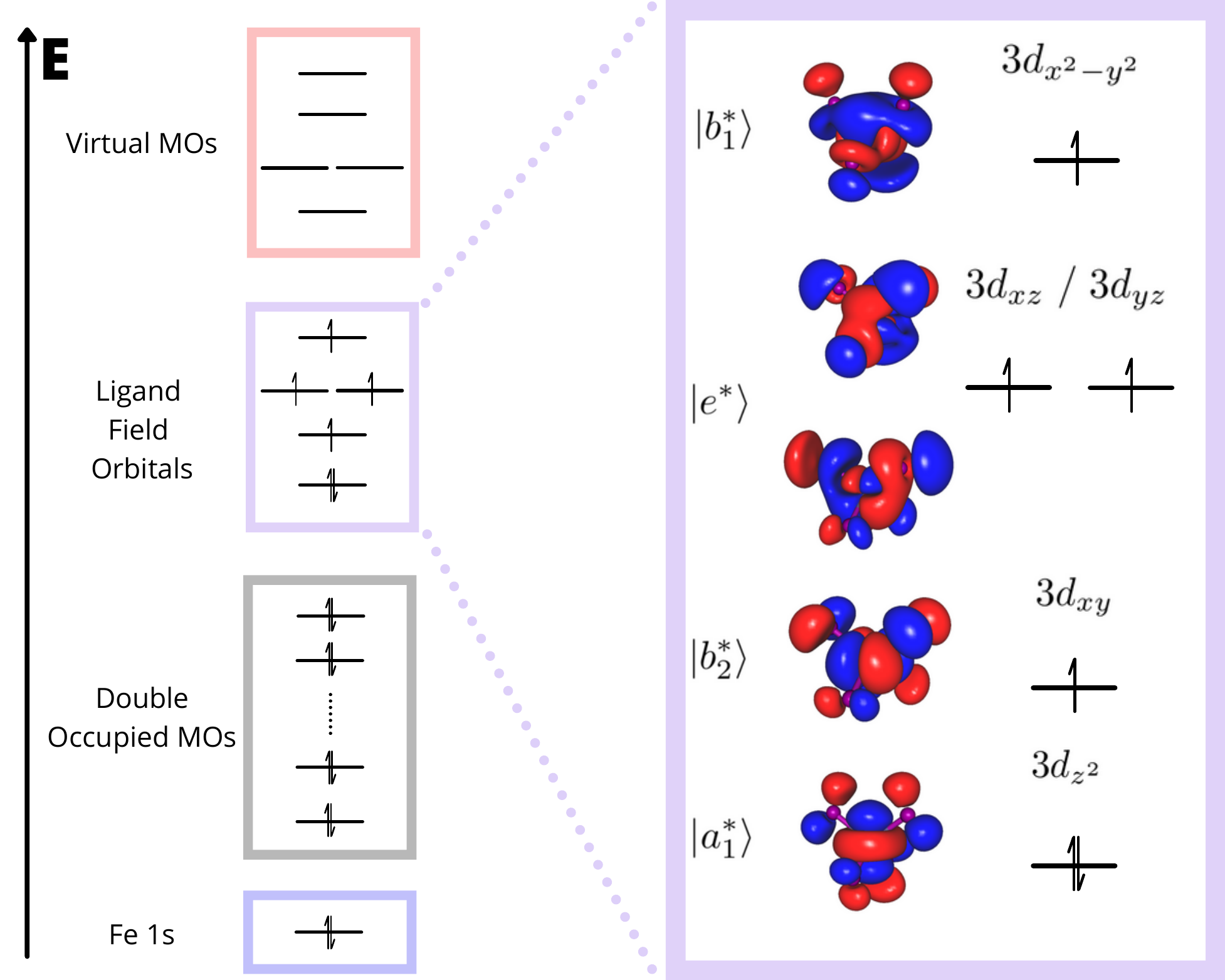}
\par\end{centering}
\caption{Left panel: the energetic ordering of fully occupied, ligand field
and virtual orbitals as obtained from our calculations. Right panel:
ligand field molecular orbitals and their ground state configuration
based on CAS(6,5)/NEVPT-2/Def2-TZVPP approach \citep{singh_covalency_2017}.
Isosurfaces values are set to $0.006$. Our results indicate a ligand
field splitting of $0.308$ eV. Ligand Orbitals are plotted via the
Gabedit program \citep{allouche_gabeditgraphical_2011} \label{fig:OrcaCalculations}}
\end{figure}

Being XANES a local probe to electronic structure, it is reasonable
to assume that the electronic structure of a single FeAs$_{4}$ distorted
tetrahedron can be connected to the pre-edge feature which, insofar
as the Fe $K$-edge is concerned, is the main focus of our paper.
Our results are presented in figure \ref{fig:OrcaCalculations} and
in table \ref{tab:orbitalcompo}. In doing that, we are assuming that
the XANES transitions in our materials are mainly due to local processes,
which excludes the metal-metal charge transfer that is observed in
certain transition metal complexes \citep{gougoussis_intrinsic_2009,cabaret_first-principles_2010}.
One can justify this assumption based on the As chemistry, which tends
to form complexes with a low bridging character, with the As $4p$
states well-localized on the As sites. This orbital localization would
weaken the Fe-Fe charge transfer that relies on the mediation of the
ligand orbitals.

The multiconfigurational calculations results show that the first
set of partially occupied states are indeed ligand field (or crystal
field) levels formed by hybrid orbitals from Fe$3d$ and As$4p$ states.
In our discussion, we name the $\mathcal{j}a_{1}^{*}\mathcal{i}$,
$\mathcal{j}b_{2}^{*}\mathcal{i}$, $\mathcal{j}e^{*}\mathcal{i}$
and $\mathcal{j}b_{1}^{*}\mathcal{i}$ molecular orbitals by their
Fe $3d$ main character, respectively, $d_{z^{2}}$,$d_{xy}$, $d_{xz}/d_{yz}$
and $d_{x^{2}-y^{2}}$. The calculated ligand field splitting of only
$\approx0.3$ eV is in qualitative agreement with previous calculations
\citep{haule_coherenceincoherence_2009}.

From table \ref{tab:orbitalcompo} and from the isosurface plot of
the molecular orbital wavefunctions in figure \ref{fig:OrcaCalculations},
one can observe that all $d_{xy}$, $d_{xz}/d_{yz}$ and $d_{x^{2}-y^{2}}$
hybridize with the As $4p_{x,y,z}$ orbitals. The $d_{z^{2}}$ and
$d_{xy}$ orbitals appear in our calculations as nearly degenerate
states and are indicated to be virtually double occupied. Electronic
transitions are thus dominated by the $d_{xz}/d_{yz}$ and $d_{x^{2}-y^{2}}$
hybrid orbitals.

All the $d_{xz}/d_{yz}$ and $d_{x^{2}-y^{2}}$ display contributions
with $p_{z}$ orbital character and therefore contribute to the increase
of the spectra intensity for $\phi$ rotations. It should be noted
that even if the $d_{xy}$ orbital were not double occupied, it would
not contribute to this, since they present no mixing with $p_{z}$
orbitals. The calculated hybridization pattern of the ligand field
orbitals is imposed by the specific properties of the FeAs$_{4}$
complex. Indeed, the $D_{2d}$ symmetry would allow the mixing of
$p_{z}$ orbitals into the $d_{xy}$ orbital, whereas it would prevent
any $p_{z}$ orbital mixing with $d_{x^{2}-y^{2}}$.

The most prominent feature to be observed from our calculations is
the contribution from the Fe$4p$ states to ligand field orbitals,
forming Fe$3d4p$ hybrid states. This local $pd$ hybridization is
acquired by the Fe$3d$ states as a formal way of reducing their antibonding
character, which is implied by the $\mathrm{As^{-3}}$ $\pi$-donation.
This effect is a mechanism for the localization of the Fe$3d$ states,
as observed in other coordination complexes \citep{alvarez_how_2006}.
As a consequence, the pre-edge peak in the Fe $K$-edge can be attributed
to Fe$3d4p$ hybridization in Fe complexes \citep{hocking_ligand_2012}
and our findings propose that we should reconsider the nature of the
pre-edge transition of the Fe $K$-edge of the FePns materials. Indeed,
so far in our discussion, as well as in previous works \citep{lafuerza_evidence_2017,pelliciari_fluctuating_2021,chen_x-ray_2011,chen_x-ray_2011-1,chang_angular_2009},
the contribution of the Fe$4p$ states to the pre-edge was overlooked.
We shall argue that this local hybridization is key to understand
the effects of Mn substitution.

First, we expect that via the Fe$3d$As$4p$ mixing, the Fe $K$-edge
results should mirror that of the As $K$-edge, as observed. This
expectation, however, does not take into account that the pre-edge
intensity may be dominated by transitions to the Fe$3d4p$ states.
This is likely the case of FeSe materials \citep{joseph_study_2010},
for which the Se $K$-edge clearly indicates that the Se$4p_{x,y}$
planar orbitals dominate the density of unoccupied states, whereas
the contrary is concluded from the Fe $K$-edge.

Mn substitution may weak the Fe$3d$As$4p$ mixing, making the Fe
orbitals more localized through the mechanisms above explained. In
turn, this would make the Fe$3d4p_{z}$ mixing stronger, increasing
the observed polarization dependence. In the same direction, the As
$4p$ will also become more localized, rendering the As $K$ edge
spectra more anisotropic. Both effects are observed in figures \ref{fig:FeKedgeplus}$(b)$
and \ref{fig:AsKedgeplus}$(b)$ and we propose that this is the mechanism
related to Mn substitution in BaFe$_{2}$As$_{2}$: Mn impurities
localize the Fe$3d$ states by changing the Fe$3d$As$4p$ mixing.
In turn, it shows that hole doping is not the only active mechanism
pushing BaFe$_{2}$As$_{2}$ to a more correlated Mott phase.

\begin{table*}
\begin{centering}
\caption{Metal and ligand orbital composition of the FeAs$_{4}$ ligand field
molecular orbitals. Numbers are given in percentage of normalized
wavefunctions. The Final orbital composition analysis was performed
adopting the Ros-Schuit partition method via the Multifwn program
\citep{lu_multiwfn_2012}. \label{tab:orbitalcompo}}
\par\end{centering}
\centering{}%
\begin{tabular}{cccccc}
\hline 
\noalign{\vskip\doublerulesep}
 & Iron orbitals (MO symmetry) & $3d_{z^{2}}$ $(a_{1}^{*})$ & $3d_{xy}$ $(b_{2}^{*})$ & $3d_{xz},3d_{yz}$ $(e^{*})$ & $3d_{x^{2}-y^{2}}$ $(b_{1}^{*})$\tabularnewline[\doublerulesep]
\noalign{\vskip\doublerulesep}
\noalign{\vskip\doublerulesep}
 & Ligand Field Relative energies / eV & 0.000 & 0.002 & 0.260 & 0.308\tabularnewline[\doublerulesep]
\hline 
\noalign{\vskip\doublerulesep}
\noalign{\vskip\doublerulesep}
\multirow{2}{*}{Fe orbitals} & $4p_{z}$ &  &  & 4.48 & 2.25\tabularnewline[\doublerulesep]
\noalign{\vskip\doublerulesep}
\noalign{\vskip\doublerulesep}
 & $4p_{y}$ and $4p_{x}$ & 5.80 &  & 10.32 & 5.80\tabularnewline[\doublerulesep]
\cline{2-6} \cline{3-6} \cline{4-6} \cline{5-6} \cline{6-6} 
\noalign{\vskip\doublerulesep}
\noalign{\vskip\doublerulesep}
\multirow{2}{*}{As Orbitals} & $4p_{z}$ & 2.75 &  & 2.00 & 2.38\tabularnewline[\doublerulesep]
\noalign{\vskip\doublerulesep}
\noalign{\vskip\doublerulesep}
 & $4p_{y}$ and $4p_{x}$ & 0.89 & 4.15 & 7.79 & 5.80\tabularnewline[\doublerulesep]
\hline 
\noalign{\vskip\doublerulesep}
\end{tabular}
\end{table*}

\section{Summary and Conclusions}

We have investigated the polarization dependence of the Fe and As
$K$-edges XANES spectra of BaFe$_{2}$As$_{2}$ and chemically substituted
Mn and Co materials of this parent compound. In the case of the Fe
$K$-edge, we focused our analysis on the transitions allowed by Fe$3d$As$4p$
hybrid orbitals that spam the pre-edge structure of the Fe $K$-edge
spectra. In the case of the As $K$-edge, we focused on the edge transitions,
probing the As $4p_{x,y,z}$ orbitals.

The polarization dependence indicates that Co substitution populates
preferentially the hybrid bands with unoccupied states along the $c$
axis ($p_{z}$ orbitals), as concluded from the strong reduction of
the As $K$-edge anisotropy. This is a distinct anisotropic charge
transfer effect, which may be connected to the transport properties
and ARPES experiments of Co doped BaFe$_{2}$As$_{2}$ \citep{nakajima_comprehensive_2018,tanatar_systematics_2011,thirupathaiah_orbital_2010}.

Mn substitution, whereas not changing the material electronic filling,
increases the anisotropy of the probed electronic states. We attributed
this finding to a delicate interplay between the local Fe$3d4p$ and
the metal-ligand Fe$3d$As$4p$ mixings, with Mn substitution favoring
the Fe$3d$ localization by hindering the Fe$3d$As$4p$ mixing.

In all cases, the XANES polarization dependence revealed a higher
density of unoccupied state for orbitals with $p_{z}$ character,
with the results from the Fe $K$-edge mirroring those of the pnictide
$K$-edge. Our quantum chemistry calculations show this is not the
only result that could be expected, since the local Fe$3d4p$ hybridization
also contributes to the pre-edge transitions and may dominate its
polarization dependence. This is likely the case of FeSe materials
\citep{joseph_study_2010}. One can thus state a clear distinct behavior
of the electronic states of iron arsenides and selenides, with the
former presenting a stronger Fe$3d$As$4p$ hybridization, which favors
the occupation of orbitals with planar geometry, whereas orbitals
along the $c$-axis remain unoccupied.

Overall, our findings suggest that the interplay between the local
Fe$3d4p$ and the metal-ligand Fe$3d$As(Se)$4p$ mixings is a common
thread of the FePns electronic structure, unveiling the key role played
by Fe $4p$ states.

\section{Acknowledgments}

The authors acknowledge CNPEM-LNLS for the concession of beamtime
at the XDS beamline (proposals Nos. 20180194 and 20190123). The XDS
beamline staff is acknowledged for their assistance during the experiments.
The Fundação de Amparo à Pesquisa do Estado de São Paulo (FAPESP)
financial support is acknowledged by M.R.C. (\#2019/05150-7 and \#2020/13701-0)
W.R.S.N. (\#2019/23879-4), D.S.C. (\#2019/04196-3), J.C.S. (\#2018/11364-7
and \#2020/12283-0), M.M.P. (\#2015/15665-3), P.G.P. and C.A. (\#2017/10581-1)
and F.A.G. (\#2019/25665-1). P.G.P. and C.A. acknowledge the financial
support from CNPq: CNPq \# 304496/2017-0 and CNPq \#310373/2019-0.

\bibliographystyle{apsrev4-2}
\bibliography{2021AdeFiguereido_References_XASBFACo-Mn_Asorbitals}

\end{document}